\documentstyle[12pt,eclepsf]{article}
\newcommand{\postscript}[1]{\hbox{\epsfile{file=#1}}}
\begin{document}
\begin{center}
\begin{Large}
{\bf Complex step-size dependences }\\
       {\bf in tracking a simple two-body dynamics}
 \footnote{RIKEN Review {\bf No. 15}(1997)95.}
 \\
\end{Large}
\end{center}
\begin{center}
    Ken Umeno\\
   {\it Laboratory for Information Representation}\\
   {\it Frontier Research Program} 
  {\it The Institute of Physical and Chemical Research (RIKEN)} 
\end{center}
\par
\vspace{0.2 cm}
\begin{small}
\begin{center}
$\hspace{-0.65 cm}$ ${\bf Abstract}$ $\hspace{0.5 cm}$
\end{center}
In molecular simulations, one of the most difficult points is to track
 the real dynamics of many-body systems from the first principle.  
The present study shows that 
step-size dependences have an unexpected effect on simulation results, 
 even when we use 
the standard high-precision numerical integrators 
to apply to a simple system with a two-body interaction.
The validity of our analysis is checked by    
the theory of adiabatic approximations.

\end{small}
\normalsize

Recently, various numerical integration techniques are developed to simulate 
the dynamics of complex systems with many-body interactions. 
Above all,  {\it symplectic integrators} are important because 
they exactly  preserve the symplectic form
      \(\omega^{2}=d\mbox{\boldmath$p$}\wedge d\mbox{\boldmath$q$}=
     \sum_{i=1,n}dp_{i}\wedge dq_{i}\)
which any natural  Hamiltonian \(H(\mbox{\boldmath$q,p$})\) must have.
Thus, it is very important to investigate and estimate how these 
symplectic integrators improve our capacity of tracking many-body dynamics.
We consider the following simple  system to be tested:
\begin{equation}
\label{eq:hami}
      H=\frac{1}{2}(p_{1}^{2}+p_{2}^{2}+q_{1}^{2}q_{2}^{2}).
\end{equation}
The equations of motion remain invariant
 under the transformation \((t'\rightarrow \alpha t,\mbox{\boldmath$q'$}\rightarrow \frac{1}{\alpha}
\mbox{\boldmath$q$},\mbox{\boldmath$p'$}\rightarrow \frac{1}{\alpha^{2}}\mbox{\boldmath$p$}\)), where
\(t\) is a time variable and \(\alpha\) 
is an arbitrary real and dimensionless quantity which does not vanish.
Thus, our simulations  do not depend on the scale of time.  
It is known that this quartic potential system shows chaotic behavior, as 
is predicted by  the non-integrability proof\cite{zig}. 
 Furthermore, the system (\ref{eq:hami}) allows not only chaotic but also  intermittent 
behavior, which significantly affects  the accuracy of 
the higher-order symplectic simulations\cite{ku,suku}. 

In this paper, 
 we investigate the time step-size dependences of its  
 simulations by two different {\em explicit} type higher-order
 symplectic integrators, namely 
 (i) Suzuki's third-order symplectic integrator 
 and (ii) Ruth's third-order symplectic integrator, 
which are 
obtained by the real 
 decompositions of exponential operators;
\begin{equation}
 e^{\left(A+B\right)\Delta t}= e^{c_{1}A\Delta t}e^{d_{1}B\Delta t}
e^{c_{2}A\Delta t}e^{d_{2}B\Delta t}e^{c_{3}A\Delta t}e^{d_{3}B\Delta t}
+O\left(\Delta t^{3}\right),
\end{equation}
where \(\Delta t\) is a time step-size and the coefficients \(c_{1},c_{2},c_{3},d_{1},
d_{2},d_{3}\) are given as 
\begin{eqnarray}
  c_{1}=d_{3}=0.2683300957817599\cdots\nonumber\\
  d_{1}=c_{3}=c_{1}+0.651331427356399\cdots\nonumber\\
  c_{2}=d_{2}=d_{1}-0.839230460347997\cdots,
\end{eqnarray}
in (i)Suzuki's case\cite{su} or they are given as 
\begin{equation}
\label{eq:third}
  c_{1}=7/24,d_{1}=2/3,c_{2}=3/4,d_{2}=-2/3,c_{3}=-1/24, d_{3}=1
\end{equation}
in (ii)Ruth's case\cite{ru}. 
The main reason why we use third-order symplectic integrators here lies
in the fact that 
they are already higher-order than the popular leap-flog method
(the second-order integrator) and it is also guaranteed  that there
are no other  third-order symplectic 
integrators with real coefficients\cite{su}.
In cases of separable Hamiltonian systems  
   \( H(\mbox{\boldmath$q,p$})=K(\mbox{\boldmath$p$})
   +V(\mbox{\boldmath$q$})\),
the implementation of these third-order symplectic  
integrators is straightforward as follows:
\begin{eqnarray}
\label{eq:split}
  \mbox{\boldmath$p$}^{(1)}=\mbox{\boldmath$p$}^{(0)}
    -c_{1}\Delta t V_{\mbox{\boldmath$q$}}(\mbox{\boldmath$q$}^{(0)}),\quad
  \mbox{\boldmath$q$}^{(1)}=\mbox{\boldmath$q$}^{(0)}+
d_{1}\Delta t K_{\mbox{\boldmath$p$}}(\mbox{\boldmath$p$}^{(1)}),\nonumber\\
  \mbox{\boldmath$p$}^{(2)}=\mbox{\boldmath$p$}^{(1)}
 -c_{2}\Delta t V_{\mbox{\boldmath$q$}}(\mbox{\boldmath$q$}^{(1)}),\quad
  \mbox{\boldmath$q$}^{(2)}=\mbox{\boldmath$q$}^{(1)}+
 d_{2}\Delta t K_{\mbox{\boldmath$p$}}(\mbox{\boldmath$p$}^{(2)}),\nonumber\\
  \mbox{\boldmath$p$}^{(3)}=\mbox{\boldmath$p$}^{(2)}-
  c_{3}\Delta t V_{\mbox{\boldmath$q$}}(\mbox{\boldmath$q$}^{(2)}),\quad
 \mbox{\boldmath$q$}^{(3)}=\mbox{\boldmath$q$}^{(2)}+
d_{3}\Delta t K_{\mbox{\boldmath$p$}}(\mbox{\boldmath$p$}^{(3)}),
\end{eqnarray}
 where 
\(  \frac{\partial}{\partial\mbox{\boldmath$q$}}V(\mbox{\boldmath$q$})
  =V_{\mbox{\boldmath$q$}}(\mbox{\boldmath$q$}),\quad
  \frac{\partial}{\partial\mbox{\boldmath$p$}}K(\mbox{\boldmath$p$})
 =K_{\mbox{\boldmath$p$}}(\mbox{\boldmath$p$})\).
In our simulations here, we always fix the initial condition as 
\((q_{1},q_{2},p_{1},p_{2})=(1000,0.002,0,0)\) to cause a strong intermittency .  
Around this initial condition, we can employ 
the adiabatic approximation as 
\begin{equation}
  H=\frac{1}{2}p^{2}_{1}+(\frac{1}{2}p^{2}_{2}+\frac{1}{2}kq^{2}_{2})=E,
\end{equation}
 where \(q^{2}_{1}\equiv k\approx \mbox{Const.}\) 
is a slow variable. A fast dynamics is simply
described
by the harmonic oscillator     
  \(H_{F}=\frac{1}{2}p^{2}_{2}+\frac{1}{2}kq^{2}_{2}=E_{F}\)
 with the 
spring coefficient \(k=q^{2}_{1}\).
Thus, the 
adiabatic invariant
\(J\) is given by the formula
\begin{equation}
  J=\sqrt{E_{F}^{2}/k}=\frac{\frac{1}{2}p^{2}_{2}+\frac{1}{2}q^{2}_{1}
q^{2}_{2}}{|q_{1}|}=
\sqrt{4/1000^{2}}=0.002,
\end{equation}
because 
\(k=q^{2}_{1}=1000^{2}\) and \(E^{2}=4\) hold 
at the initial condition. 
Thus, we have a slow dynamics given by the Hamiltonian  
  \(H=\frac{1}{2}p^{2}_{1}+J\cdot|q_{1}|\) 
and therefore we can obtain the analytic solution of this slow dynamics 
\begin{equation}
  q_{1}(t)=q_{1}(t=0)-\frac{1}{2}Jt^{2}.
\end{equation} 
By considering that the hypothesis of this adiabatic approximation is based on the assumption 
\(|q_{1}|\gg 1\), an {\it adiabatic transition} is predicted to occur
  at \(t=T\) given by 
\begin{equation}
\label{eq:formula}
 T=\sqrt{2q_{1}(t=0)/J}=\sqrt{2\times 1000/0.002}=1000,
\end{equation}
where \(q_{1}(t=T)\approx 0 \). 
Figure 1 shows that the third-order symplectic integrator with a step-size \(\Delta t=0.001\) 
 can correctly track the adiabatic transition at \(t=1000\). 
 Figure 2 shows that a slight 
step-size difference cause the great discrepancy in tracking   
 the trajectory
  \((q_{1}(t),q_{2}(t))\), even though  each third-order 
 symplectic integrator with  the step-size around \(\Delta t=0.001\) 
 can give the correct behavior of adiabatic transitions like 
 Fig.1. 
Figure 3.(a)-(c) show that 
this effect of step-size dependences of the dynamical variable \(q_{2}\) 
at the fixed transition time \(t=1000\) is extraordinary  complex 
beyond our imagination. Remark that this effect does not 
depend on the choice of integrators as is indicated in Fig.3.(a)-(c).
This result suggests that the following usual 
argument that  ``if \(\Delta t=0.002\) is not O.K, then \(\Delta t=0.001\) is 
maybe O.K.'' is generally untrue unless 
the step-size \(\Delta t\) is sufficiently  small such as
 \(0<\Delta t\leq 0.0001\) like 
 Fig.3.(c). 
Thus, this kind of complex behavior of step-size dependences can be 
a real challenge to the  solid progress in the studies of many-body dynamics simulations, since 
the intractability here 
is closely related to the non-integrable character of generic many-body 
systems\cite{ku1,ku2},  
whether the simulations are classical or quantum ones\cite{ku3}. 
\clearpage
{\bf Figures}\\

\begin{figure}[htb]
\begin{minipage}[t]{8.0cm}
\postscript{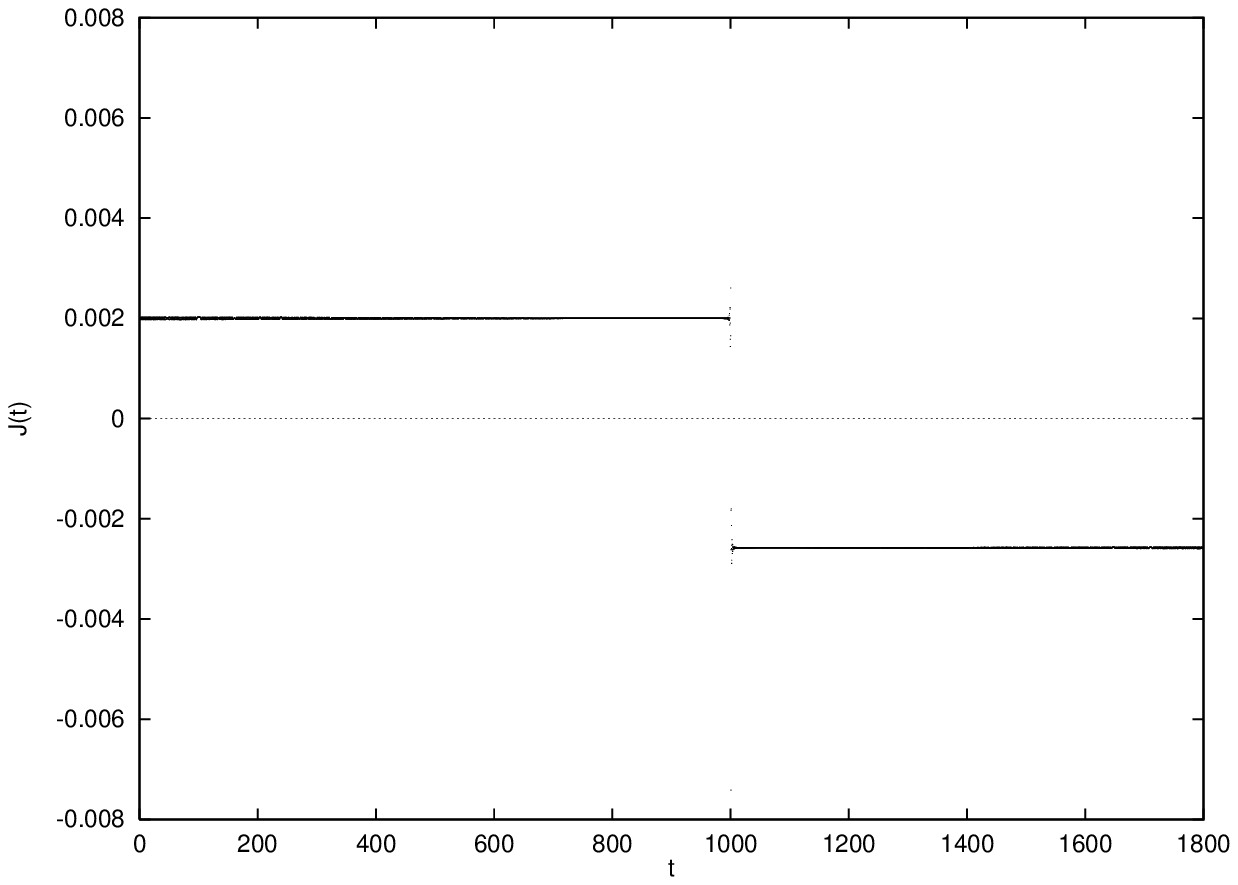,height=10cm}
\end{minipage} \hfill \\
{{\bf Fig. 1:}The time evolution of the adiabatic invariant \( J(t) \) 
calculated 
using the third-order symplectic integrator for \(\Delta t=0.001\) with 
the initial condition \(q_{1}=1000,q_{2}=0.002\) and \(p_{1}=p_{2}=0\).
An adiabatic transition 
is observed just at  \(t=1000\), as is predicted in Eq. (\ref{eq:formula}).  
}
\end{figure}

\begin{figure}[htb]
\begin{minipage}[t]{8.0cm}
\postscript{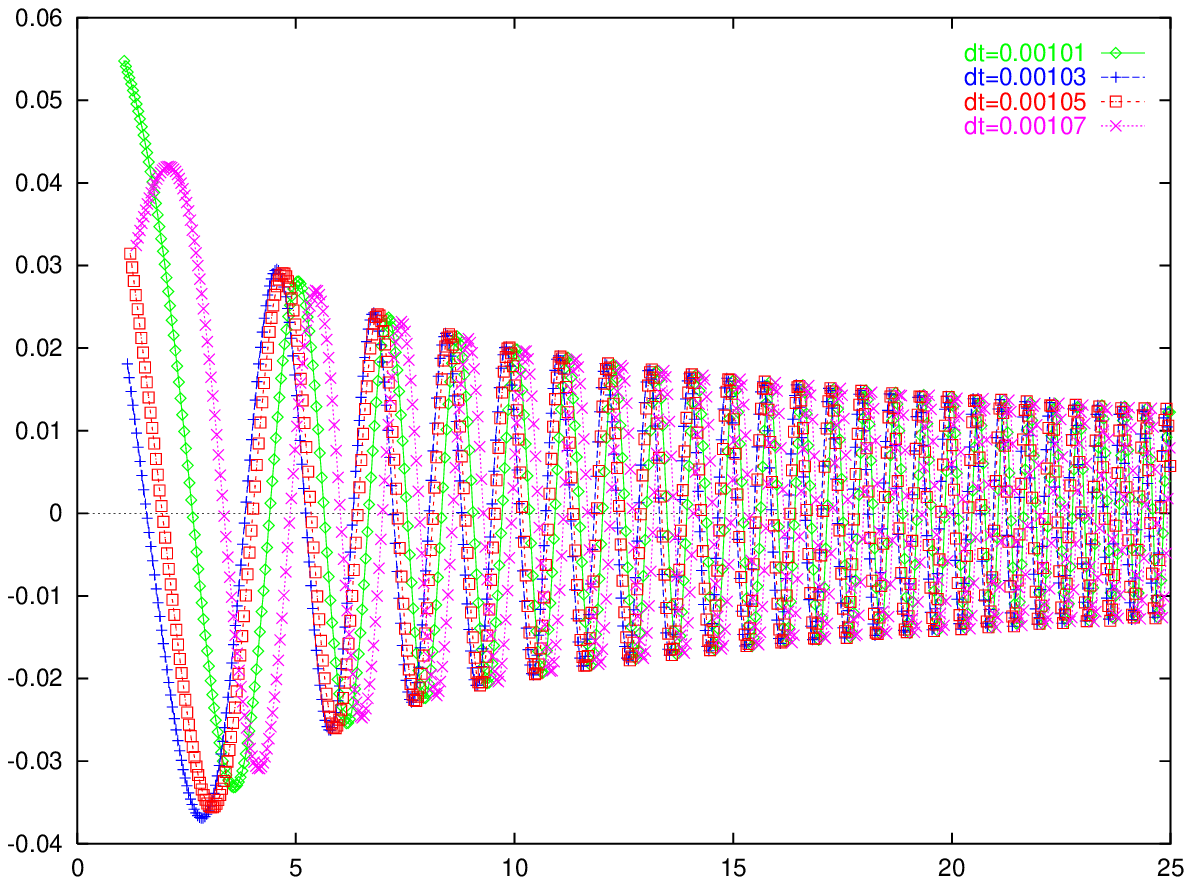,height=10cm}
\end{minipage} \hfill \\
{{\bf Fig. 2:}
The trajectories of \((q_{1},q_{2})\) calculated using the third-order 
 symplectic integrator for \(\Delta t=0.00101\)(Green), \(\Delta t=0.00103\)
(Blue), \(\Delta t=0.00105\)(Red), and \(\Delta t=0.00107\)(Pink) with 
the initial condition \(q_{1}=1000,q_{2}=0.002\) and \(p_{1}=p_{2}=0\).
}
\end{figure}

\begin{figure}[htb]
\begin{minipage}[t]{8.0cm}
\postscript{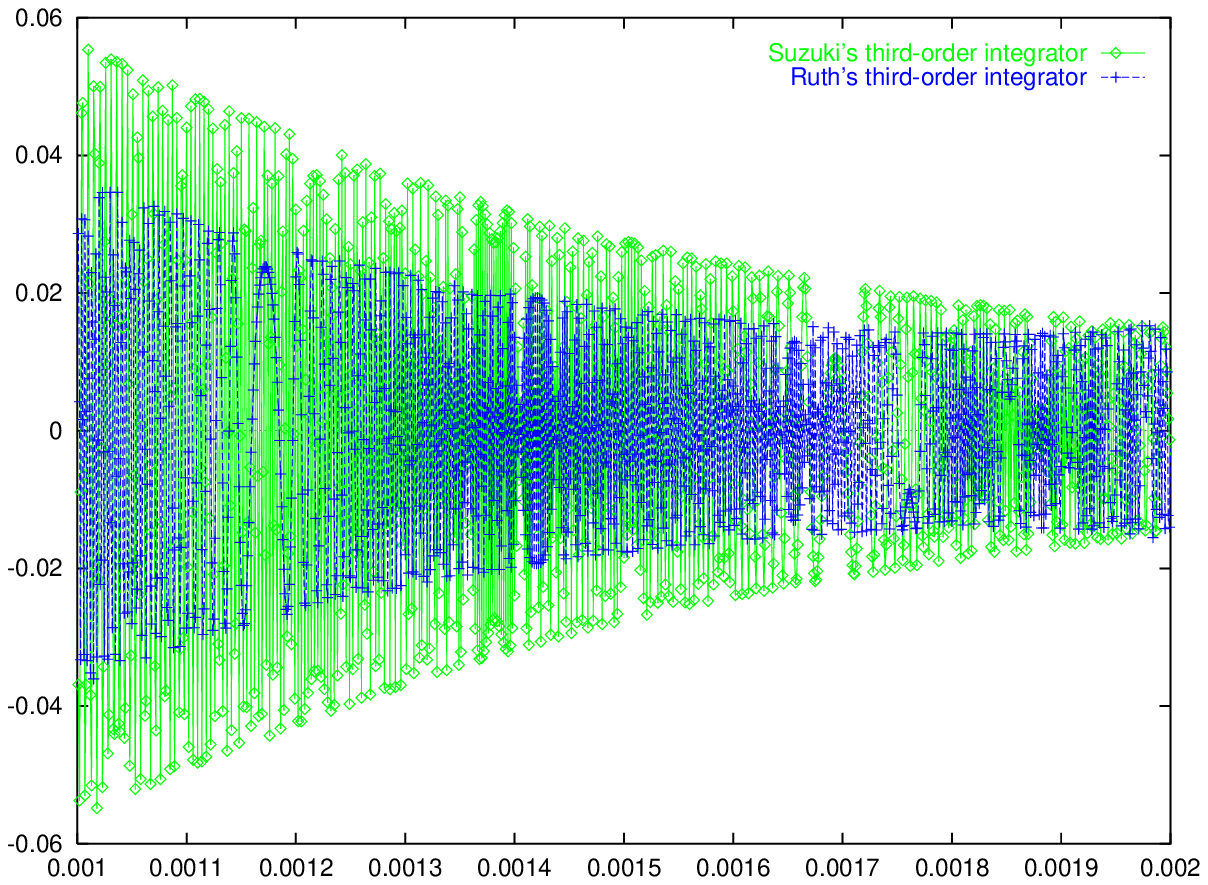,height=10cm}
\end{minipage} \hfill \\
{{\bf Fig. 3(a):}
The dynamical variable 
\(q_{2}\) at \(t=1000\) tracked from  
the {\it unique} initial condition 
\(q_{1}=1000,q_{2}=0.002\) and \(p_{1}=p_{2}=0\) 
at \(t=0\) using two different  third-order symplectic integrators 
(Suzuki's integrator(Green) and Ruth's integrator(Blue))  
is plotted against the  time step size \(\Delta t \)  
(\(0.001\leq \Delta t \leq 0.002\)).}
\end{figure}

\begin{figure}[htb]
\begin{minipage}[t]{8.0cm}
\postscript{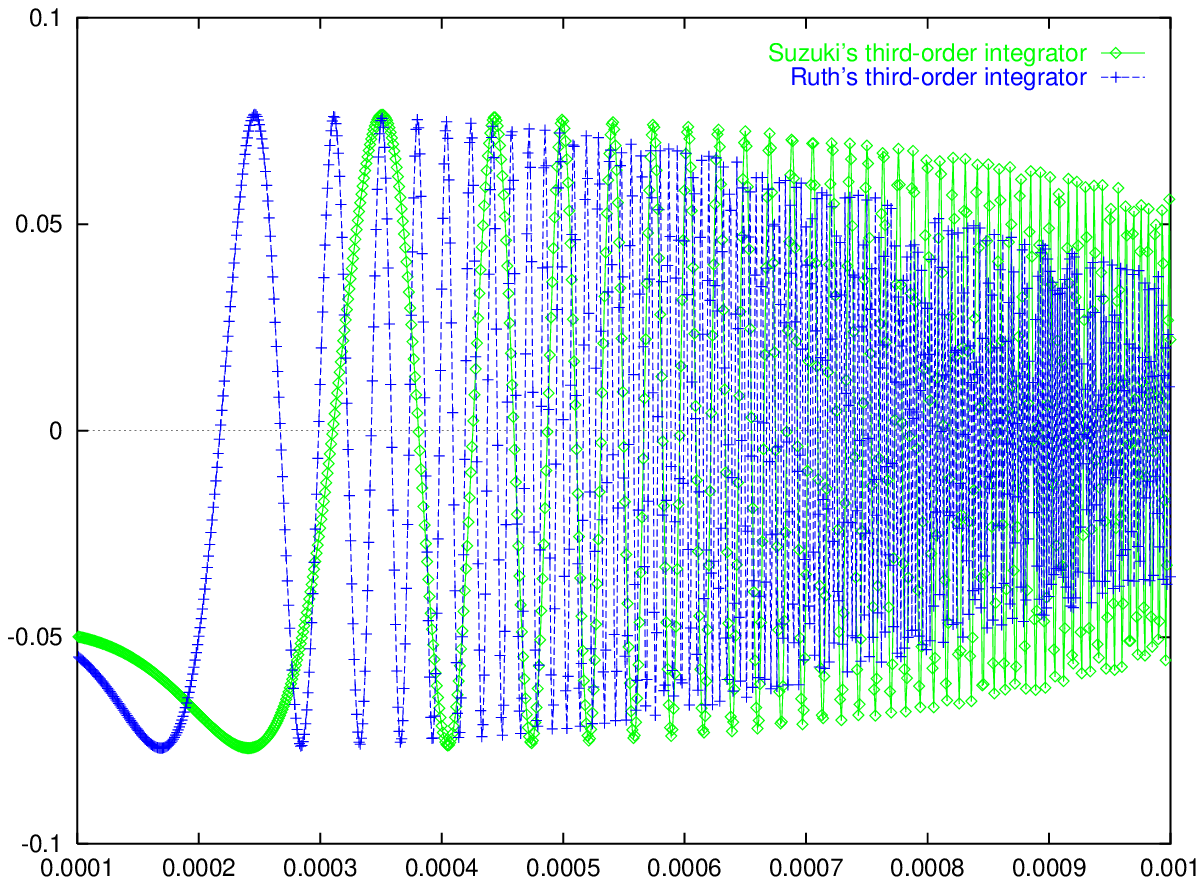,height=10cm}
\end{minipage} \hfill \\
{{\bf Fig. 3(b):}
The dynamical variable 
\(q_{2}\) at \(t=1000\) tracked from  
the {\it unique} initial condition 
\(q_{1}=1000,q_{2}=0.002\) and \(p_{1}=p_{2}=0\) 
at \(t=0\) using two different  third-order symplectic integrators 
(Suzuki's integrator(Green) and Ruth's integrator(Blue))  
is plotted against the  time step size \(\Delta t \)  
(\(0.0001\leq \Delta t \leq 0.001\)).}
\end{figure}

\begin{figure}[htb]
\begin{minipage}[t]{8.0cm}
\postscript{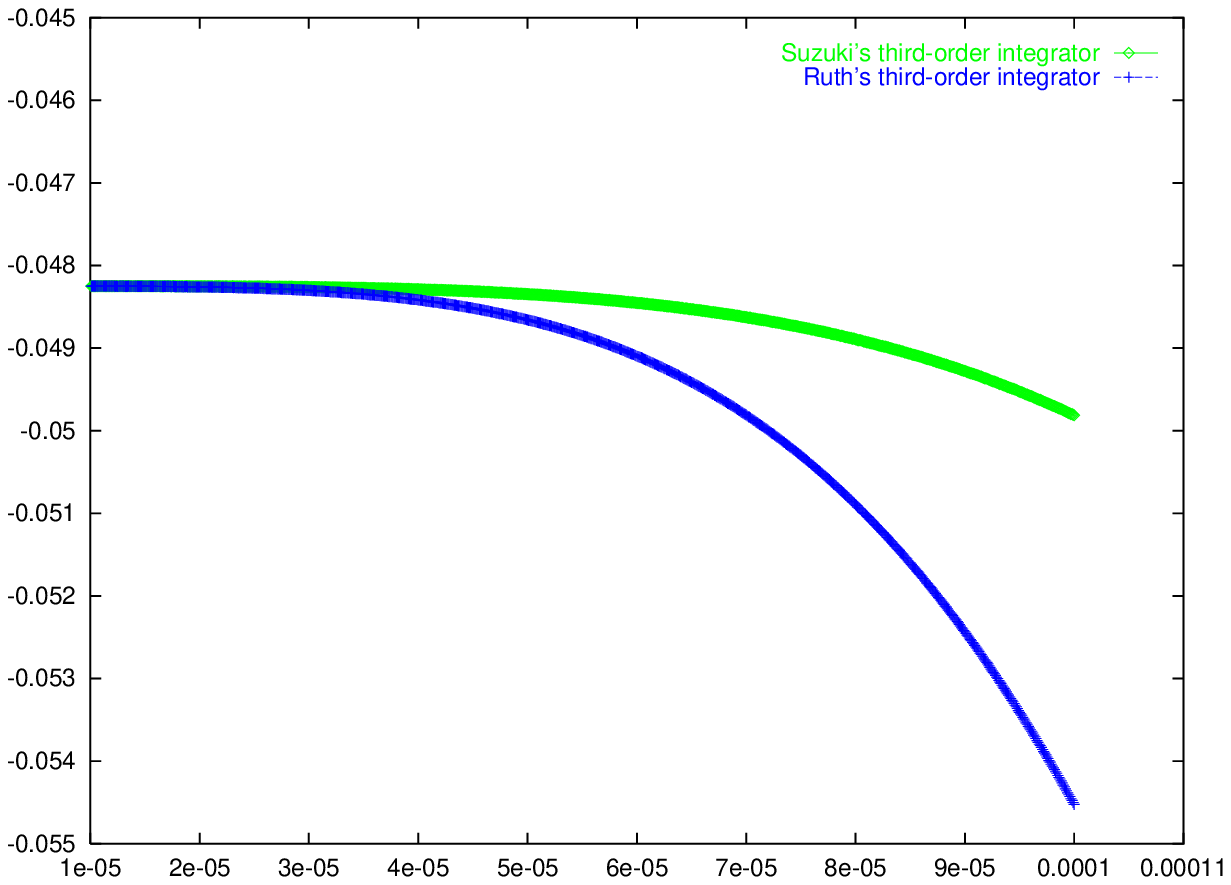,height=10cm}
\end{minipage} \hfill \\
{{\bf Fig. 3(c):}
The dynamical variable 
\(q_{2}\) at \(t=1000\) tracked from  
the {\it unique} initial condition 
\(q_{1}=1000,q_{2}=0.002\) and \(p_{1}=p_{2}=0\) 
at \(t=0\) using two different  third-order symplectic integrators 
(Suzuki's integrator(Green) and Ruth's integrator(Blue))  
is plotted against the  time step size \(\Delta t \)  
(\(0.00001 \leq \Delta t 
\leq  0.0001\)). 
About \(\frac{1000}{\Delta t} \) iterations are required to estimate  
each \(q_{2}(t=1000;\Delta t)\).   
}
\end{figure}
\end{document}